\documentclass[]{aa}  

\usepackage{graphicx}
\usepackage{natbib}
\usepackage{amssymb}
\usepackage{amsmath}
\usepackage{xspace}
\usepackage{color}
\usepackage[normalem]{ulem}
\usepackage[varg]{txfonts}
\newcommand{\be}{\begin{equation}}
\newcommand{\ee}{\end{equation}}

\newcommand{\rb}[1]{{ \color{black}  #1}}

\begin{document}

   \title{Non-thermal emission from the reverse shock of the youngest galactic Supernova remnant G1.9+0.3}

   \author{R. Brose\inst{1,2}\fnmsep\thanks{Corresponding author, \email{robert.brose@desy.de}}
          \and
	  I. Sushch\inst{1,3,7}
          \and
      M. Pohl\inst{1,2}
	 	 \and
	  K. J. Luken\inst{4}
	  	  \and
	  M. D. Filipovi\'{c}\inst{4}
	  	  \and
	  R. Lin\inst{5,6}}

\institute{DESY, 15738 Zeuthen, Germany 
\and Institute of Physics and Astronomy, University of Potsdam, 14476 Potsdam, Germany
\and Astronomical Observatory of Ivan Franko National University of L'viv, vul. Kyryla i Methodia, 8, L'viv 79005, Ukraine
\and Western Sydney University, Locked Bag 1797, Penrith NSW 2751, Australia
\and Sorbonne University, 4 place Jussieu, 75005 Paris
\and Universit\'{e} de Versailles Saint-Quentin-en-Yvelines, 45 rue des \'{E}tats Unis, 78000 Versailles
\and Centre for Space Research, North-West University, 2520 Potcheftroom, South Africa }

\date{Received ; accepted }


  \abstract
   {The youngest Galactic supernova remnant G1.9+0.3 is an interesting target for next generation gamma-ray observatories. So far, the remnant is only detected in the radio and the X-ray bands, but its young age of $\approx 100\,$yrs and inferred shock speed of $\approx14,000\,$km/s could make it an efficient particle accelerator.}  
   {We aim to model the observed radio and X-ray spectra together with the morphology of the remnant. At the same time, we aim to estimate the gamma-ray flux from the source and evaluated the prospects of its detection with future gamma-ray experiments.}
  {We performed spherical symmetric 1-D simulations with the RATPaC code, in which we simultaneously solve the transport equation for cosmic rays, the transport equation for magnetic turbulence, and the hydro-dynamical equations for the gas flow. Separately computed distributions of the particles accelerated at the forward and the reverse shock are then used to calculate the spectra of synchrotron, inverse Compton, and pion-decay radiation from the source.}
  {The emission from G1.9+0.3 can be self-consistently explained within the test-particle limit. We find that the X-ray flux is dominated by emission from the forward shock while most of the radio emission originates near the reverse shock, which makes G1.9+0.3 the first remnant with non-thermal radiation detected from the reverse shock. The flux of very-high-energy gamma-ray emission from G1.9+0.3 is expected to be close to the sensitivity threshold of the Cherenkov Telescope Array, CTA. The limited time available to grow large-scale turbulence limits the maximum energy of particles to values below 100$\,$TeV, hence G1.9+0.3 is not a PeVatron.}
   {}


   \maketitle
%

\section{Introduction}
The approximately $100\,$-yr old supernova remnant (SNR) G1.9+0.3 is presumably the youngest SNR in our Galaxy. It was first detected in the radio band by \cite{1984Natur.312..527G} and in X-rays by \cite{2008ApJ...680L..41R}. The measured gas column density places the remnant close to the Galactic center at about $8.5\,$kpc and strong absorption in the optical {band} led to the non-detection of the supernova explosion. The seemingly circular shell has a radius of about $1.75\,$pc and an inferred expansion velocity of about $11,000\,$km/s at this distance \citep{2008MNRAS.387L..54G}.

The radio shell of G1.9+0.3 shows a strong north-south (N-S) asymmetry with the brighter rim towards the north and the dimmer rim towards the south. Radio measurements showed a brightening of G1.9+0.3 with an annual flux increase of about $1.2\,$\%/yr$^{-1}$ \citep{2008MNRAS.389L..23M} that seems to be ongoing (Luken et al. in prep.). Radio observations of G1.9+0.3 now span roughly one third of the SNRs lifespan, promising unique insights into the underlying processes.

The X-ray structure of G1.9+0.3 consists of a main shell with a radius of about $2\,$pc and symmetric extensions on the E and W sides that extend to about $2.2\,$pc \citep{2008ApJ...680L..41R}. Proper-motion measurements of the X-ray features revealed a highly anisotropic expansion of the remnant with the fastest expansion in the E-W direction at speeds up to $15,000$km/s \citep{2017ApJ...837L...7B}. Thermal emission from intermediate-mass elements and iron reveals an asymmetric distribution with prominent Fe K$\alpha$ emission from the northern, radio-bright rim \citep{2013ApJ...771L...9B}. The asymmetric distribution of ejecta hints at a highly asymmetric explosion, which could explain the radio N-S asymmetry. A gradient in the density of the ambient medium would be an alternative explanation.

So far no detection of gamma-ray emission has been reported by H.E.S.S. or Fermi \citep{2014MNRAS.441..790H} which constrains the {average magnetic-field strength} {inside the remnant} to $B>17\,\mu$G.

\cite{2017A&A...603A...7A} analyzed the acceleration efficiency of particles in G1.9+0.3 using the synchrotron emission and found that it is only 5\% of the efficiency expected for Bohm-like diffusion, whereas values close to 100\% were deduced for the slightly older SNRs Cas A \citep{2014ApJ...785..130Z} and
RX J1713.4-3946 \citep{2010ApJ...708..965Z}.

Modeling of the particle acceleration in G1.9+0.3 by \cite{2017MNRAS.468.1616P} showed that the spectral-energy distribution and the brightening-rate of the remnant can be explained using non-linear diffusive shock acceleration (NDSA). 
However, the injection fraction of non-thermal particles entering the NDSA process must be rather high, otherwise the shock modification is too weak to reproduce the observed radio spectral index. Both published models assume that there is only one population of particles accelerated at one shock, which is in tension with the observed morphology of the remnant as the X-ray emission would peak right at the shock and the radio-emission would peak slightly inward by a few \% of the shock radius, $R_s$. The observed separation of the two intensity peaks is on the order of 0.25~$R_\mathrm{sh}$ \citep{2008ApJ...680L..41R}. 

Efficient particle acceleration to PeV-energies might only take place during the very early stages of SNR evolution \citep[and references therein]{2013MNRAS.431..415B}. The very young age of G1.9+0.3 combined with the high expansion velocity of about $14,000\,$km/s make this Galactic SNR a prime PeVatron candidate and a primary target for observations of the Cherenkov Telescope Array (CTA) \citep{2017arXiv170907997C}. 

The aim of this paper is to model the broadband emission of G1.9+0.3 and to explain the morphology and the observed radio-brightening. To do so we combine high-resolution hydrodynamical simulations based on the Pluto code \citep{2007ApJS..170..228M} with time-dependent 1D simulations of the particle acceleration at forward and reverse shock of the SNR, as well as the transport of magnetic turbulence using our RATPaC (\textbf{R}adiation \textbf{A}cceleration and \textbf{T}ransport \textbf{Pa}rallel \textbf{C}ode) framework. We show that the large radial separation between the intensity peaks of radio and X-ray emission indicates that a sizable fraction of the emission originates at the reverse shock, as expected for young SNRs \citep{Telezhinsky.2012a}.

In Section \ref{sec:pa_ac}, \ref{sec:turb_tran} and \ref{sec:hd_tran} we will describe our treatment of particle acceleration, magnetic turbulence, and hydrodynamical flow profiles. In Section \ref{sec:one-shock} we discuss a simple one-shock model and in Section \ref{sec:two-shock} a two-shock model including the self-consistent amplification of turbulence at the forward shock. Our findings are then summarized in Section \ref{sec:conc}.  

\section{Modeling}
\subsection{Particle acceleration}\label{sec:pa_ac}
We model the acceleration of cosmic rays using a kinetic approach in the test-particle approximation \citep{Telezhinsky.2012a,Telezhinsky.2012b,Telezhinsky.2013}, always keeping the cosmic-ray pressure below 10\% \citep{2010ApJ...721..886K} of the shock ram pressure. The time-dependent transport equation for the differential number density of cosmic rays $N$ \citep{Skilling.1975a} is given by

\begin{align}
\frac{\partial N\text{{(r,p,t)}}}{\partial t} &= \nabla(D_{\mathrm r}\nabla N-\mathbf{u} N)-\frac{\partial}{\partial p}\left( (N\dot{p})-\frac{\nabla \cdot \mathbf{ u}}{3}Np\right)+Q
\label{CRTE}\text{ , }
\end{align}
where $D_\mathrm{r}$ denotes the spatial diffusion coefficient, $\textbf{u}$ the advective velocity, $\dot{p}$ energy losses and $Q$ the source of thermal particles.

We solve this transport equation {in spherical symmetry and} in a frame co-moving with the shock. The radial coordinate is transformed according to $(x-1)=(x^*-1)^3$, where $x=r/R_\mathrm{sh}$ and $R_\mathrm{sh}$ is the shock {radius}. For an equidistant binning of $x^*$ this transformation guarantees a very fine resolution close to the shock, where $\Delta r = 10^{-6}\cdot R_\mathrm{sh}$. At the same time the outer grid-boundary extends to several tens of shock-radii upstream for $x^*>>1$. Thus all accelerated particles can be kept in the simulation domain. 

The injection of particles is determined by the source term, 
\begin{equation}
Q = \eta n_{\mathrm u}  (V_{\mathrm sh} - V_{\mathrm u}) \delta(R-R_{\mathrm sh}) \delta(p - p_{\mathrm{inj}})\ ,
\end{equation}
where $\eta$ is the injection efficiency parameter, $n_{\mathrm u}$ is the plasma number density in the upstream region, $V_{\mathrm sh}$ is the shock speed, $V_{\mathrm u}$ is the plasma velocity in the upstream region, and $p_\mathrm{inj}= \xi p_{\rm th}$ is the injection momentum, defined as a multiple of the mean thermal momentum of the plasma-particles with the temperature $T_{\rm d}$,  $p_{\rm th} = \sqrt{2mk_{\rm B}T_{\rm d}}$.
{For cosmic-ray protons} we use the parametrization of \citet{2005MNRAS.361..907B} for the thermal-leakage injection model. Here $p_{\rm inj}$ is the minimum momentum above which particles can cross the shock and participate in the acceleration process. For a strong shock with a compression ratio of 4, the injection efficiency is determined by
\begin{equation}
  \eta = \frac{4}{\sqrt{\pi}}\frac{\xi^3}{e^{\xi^2}}.
\end{equation}
We inject all particles with {momentum} $p_\mathrm{inj}$ at the position of the shock. The injection parameter $\xi$  determines the normalization of the resulting particle spectra and, in the case of a self-consistent treatment of Alfv\'{e}nic turbulence, the maximum energy of the cosmic rays (CR). If $\xi$ were the same for electrons and for protons, the resulting electron-to-proton ratio at high energies would be determined by the mass ratio, $K_{\rm ep} \simeq \sqrt{m_{\rm e}/m_{\rm p}}$ \citep{1993A&A...270...91P}. {Strictly speaking, thermal leakage can not be the only injection mechanism for electrons. The shock thickness is commensurate with the gyro radius of the incoming protons, i.e. $r_\mathrm{L}= V_{\mathrm sh}/\Omega_\mathrm{p}$, and only particles with Larmor radii a few times that see the shock as a discontinuity and can be accelerated by DSA. For typical SNR shock speeds, electrons would only get injected into the DSA process if their momentum is above a few tens of MeV/c, i.e. their momentum is larger than the thermal momentum of the protons and $\xi_\mathrm{e}>>\xi_\mathrm{p}$. However, a fraction of the electrons might be pre-accelerated at the shock, for instance by shock-surfing acceleration or shock drift acceleration \citep{2012ApJ...755..109M,PhysRevLett.119.105101, 2017ApJ...847...71B}, enabling their participation in DSA. Thermal leakage of electrons is hence a gross simplification of complicated processes but, as the relevant parameters in our model are the number of injected particles and the electron-to-proton-ratio, a detailed account of the electron pre-acceleration is not needed, and we treat the initial pre-acceleration at the shock like DSA.}

We separately solve the transport equations for the forward and reverse shock. We treat both shocks independently as there is little exchange of CRs through the contact discontinuity. The two particle distributions are then combined to calculate the total emission from the remnant.

Here, we consider two realizations of the diffusion coefficient. The first option is to assume Bohm diffusion close to the shock and in the downstream region, and to have a transition to the galactic diffusion coefficient further upstream. {In this case the diffusion coefficient is given by
\begin{align}
    D(r,p) = \zeta\frac{v}{3}r_g(r,p) = \zeta D_B(r,p)\text{ , }\label{eq:Bohm}
\end{align}
where $r_g$ is the gyro-radius of the particle, $v$ the particles velocity, $\zeta$ a free scaling parameter, and $D_B$ the diffusion coefficient in the Bohm-limit. Here, $\zeta=1$ represents the lowest diffusion coefficient possible in a given magnetic field and $\zeta>1$ might be a more common situation.} Alternatively the amplification of Alfv\'{e}nic turbulence can be {explicitely} treated by solving a separate wave transport equation and thus calculating the diffusion coefficient self-consistently \citep{2016A&A...593A..20B}. 

\subsection{Magnetic Turbulence}\label{sec:turb_tran}

There are potentially a number of wave modes that can scatter CRs inside and upstream of supernova remnants. While small-scale non-resonant modes can be produced very efficiently \citep{2000MNRAS.314...65L,2004MNRAS.353..550B}, they are not able to scatter the most energetic particles in the system via resonant interactions unless efficient mode-conversion is established. This makes it unclear whether or not these modes can support the increase of the maximum energy of the CRs in the SNR.

Alfv\'en waves have been discussed as scattering centers for CRs in SNRs for a long time. In this paper we consider isotropic Alfv\'en waves as scattering centers for CRs. The transport of magnetic turbulence can be described by a continuity equation for the spectral energy density, $E_w = E_w(r,k,t)$,
\begin{align}
 \frac{\partial E_{\rm w}{{(r,k,t)}}}{\partial t} + \mathbf{u} \cdot (\nabla E_{\rm w}) + (\nabla \cdot \mathbf{u})E_{\rm w} + k\frac{\partial}{\partial k}\left( \mathbf{k^2} D_{\rm k} \frac{\partial}{\partial k} \frac{E_{\rm w}}{k^3}\right) = \nonumber\\
=2(\Gamma_{\rm g}-\Gamma_{\rm d})E_{\rm w}
\label{Turb_1}
\text{ , }
\end{align}
where $D_k$ is the diffusion coefficient in wavenumber space representing cascading and $\Gamma_{\rm g}$ and $\Gamma_{\rm d}$ are the growth and the damping rates, respectively {\citep[and references therein]{2016A&A...593A..20B}}.

We are thus able to track the growth and damping, the cascading{, compression,} and the propagation of the Alfv\'enic turbulence in the CR precursor and inside the SNR. The transport equation for the magnetic turbulence (\ref{Turb_1}) is solved in parallel to the CR transport equation (\ref{CRTE}) for cosmic rays and the hydrodynamic equations. It is also transformed to a frame co-moving with the shock and solved in 1D {under spherical symmetry}, leading to a system of coupled equations which we numerically solve using implicit finite-difference methods provided by FiPy \citep{FiPy.2009a}.

We only consider resonant interactions between waves and CRs. The resonance condition is
\begin{align}
k_{\rm res} = \frac{qB_{\mathrm 0}}{pc} \text{ , }\label{eq:res}
\end{align}
where $k_{\rm res}$ is the wavenumber, $q$ the particle charge, and $B_0$ the amplitude of the background magnetic field.

Alfv\'en waves can be generated by particles streaming faster than the Alfv\'en speed, $v_A$. The generated waves have wavelengths similar to the gyro-radii of the particles \citep[ and references therein]{Wentzel.1974}. This mechanism, known as resonant amplification of Alfv\'en waves, is the only wave-driving process considered in this work. In the diffusion limit the growth rate of the waves can be related to the pressure gradient of the CRs. The growth rate due to resonant amplification is given as \citep{Skilling.1975a, Bell.1978a} 
\begin{align}
 \Gamma_{\rm g} &= \frac{v_{\rm A} p^2v}{3E_{\rm w}}\left|\frac{\partial N}{\partial r}\right| \label{Bell_res}\text{ , }
\end{align}
where equation (\ref{eq:res}) provides the link between $p$ and $k$.
For wave damping we consider only ion-cyclotron damping, as the young age of the remnant and the absence of a region with a low temperature makes neutral-charge collisions an inefficient damping process. The interaction of Alfv\'en \rm  is strongest at small scales,
\begin{align}
 \Gamma_{\rm d,IC} = \frac{v_{\rm A} c k^2}{2 \omega_{\rm P}}\text{ , }
\end{align}
where $\omega_{\rm P}$ is the ion plasma frequency \citep{Threlfall.2011a}. This mechanism should heat the plasma, which is not included in this work, though we are aware that it might modify the spectrum around near the injection momentum.

The energy transfer between turbulence of different scales is not yet fully understood and subject of ongoing research. If shocks are not involved, it can be empirically described as a diffusion process in wavenumber space \citep{Zhou.1990, Schlickeiser.2002a}. Given our assumption of isotropic turbulence, the corresponding diffusion coefficient is 
\begin{align}
 D_k = k^\mathbf{3} v_{\rm A}\sqrt{\frac{E_{\rm w}}{2 B_0^2}}\text{ . }
\end{align}
If cascading is the dominant process, this phenomenological treatment will result in a Kolmogorov-like turbulence spectrum, $E_{\rm w}\propto k^{-2/3}$.

The term for the amplification of turbulence in Eq.~(\ref{Turb_1}) requires some sort of seed turbulence for which we derive a spectrum from an assumed diffusion coefficient as initial condition, 
\begin{align}
 D_0 = 10^{27}\left(\frac{pc}{10\,\text{GeV}}\right)^{1/3}\left(\frac{B_0}{3\,\mu\text{G}}\right)^{-1/3}\label{D_r} \text{ . }
\end{align}
The value of $D_0$ is lower by a factor $100$ than the ISM diffusion coefficient found in studies of galactic CR propagation to improve the numerical stability of the code \citep[e.g.]{2011ApJ...729..106T}.

The amplified turbulence contributes to the overall magnetic field in the remnant and can play a crucial role in the acceleration and the emission processes. The effective magnetic-field strength, $B_\mathrm{tot}$, is given by
\begin{align}
 B_\mathrm{tot}^2 &= B_0^2+\langle \delta b^2\rangle\text{ , where }\label{eq:B_tot}\\
\langle \delta b^2\rangle &= 4\pi\int W_\mathrm{w} (k)\, dk = 4\pi \int E_\mathrm{w} (k)\, d\ln k\text{ . }
\end{align} 
We are aware that we overestimate the guide magnetic field experienced by high-energy particles and long-wavelength waves in this way. 
Both effects are of small impact and additionally act in a way to partially compensate each other.

It has been shown that additional magnetic turbulence can be generated downstream of the SNR shock by various MHD processes \citep[and references therein]{2014NIMPA.742..169F}. We included the possibility of such mechanisms by injecting large-scale turbulence downstream of the shock as a delta function in wave-number space.
\begin{align}
 Q_{\rm ls} &= \frac{\kappa}{2}\rho_{\rm d}v_{\rm d}^2\exp\left(-\frac{R_\mathrm{sh}-r}{\Delta r}\right)\delta(k-k_{\rm i}) \text{ , }
\end{align}
where {$\kappa\approx1\%$ determines the fraction of the downstream {thermal energy} that is transferred to magnetic-field energy, and $k_i=3.1\cdot10^{-17}\,$cm$^{-1}$ and $\Delta r = 0.015\cdot R_\mathrm{sh}$ have fixed values}. We inject this post-shock turbulence downstream of both shocks.

{
\subsection{Magnetic field}
We solve the induction equation for the large-scale magnetic field following \cite{Telezhinsky.2013}. The resulting magnetic field peaks at the forward shock and falls off toward the contact discontinuity. The far and immediate upstream magnetic fields are free parameters. In all cases we set $B_0=20\,\mu$G for the far-upstream magnetic field. We allow for a possible amplification of the magnetic field upstream of the forward shock that is either parametrized or derived by explicit consideration of the transport equation for magnetic turbulence superimposed on a spatially constant homogeneous magnetic field in the upstream region. The immediate upstream magnetic field gets compressed at the shock by a factor of $\sqrt{11}$.}

{For the interior of the remnant we assume that the surface magnetic field of the star is {frozen into the ejecta, leading to about $ 0.02\,\mu$G immediately} upstream of the reverse shock at the age of $140\,$yrs. 
}

\subsection{Hydrodynamical model}\label{sec:hd_tran}

We model the hydrodynamic evolution of G1.9+0.3 by solving the standard gas-dynamical equations
\begin{align}
\frac{\partial }{\partial t}\left( \begin{array}{c}
                                    \rho\\
				    \textbf{m}\\
				    E
                                   \end{array}
 \right) + \nabla\left( \begin{array}{c}
                   \rho\textbf{v}\\
		   \textbf{mv} + P\textbf{I}\\
		   (E+p)\textbf{v} 
                  \end{array}
 \right)^T &= \left(\begin{array}{c}
                    0\\
		    0\\
		    0
                   \end{array}
 \right)\\
 \frac{\rho\textbf{v}^2}{2}+\frac{P}{\gamma-1}  &= E \text{,}
\end{align}
where $\rho$ is the density of the thermal gas, $\textbf{v}$ the plasma velocity, $\textbf{m}=\textbf{v}\rho$ the momentum density, $P$ the thermal pressure, and $E$ the total energy of the ideal gas with $\gamma=5/3$. This system of equations is solved under the assumption of spherical symmetry in 1-D using the PLUTO code \citep{2007ApJS..170..228M}. We ignore the possible feedback from the CRs onto the shock structure as the ratio of the CR pressure to the shock ram pressure is always below 10\% in our simulations \citep{2010ApJ...721..886K}. {Hence} {our hydro simulations are independent from the cosmic-ray and turbulence calculations and can be used as a pre-calculated input. We used the ausm+ solver implemented in PLUTO and a CFL-value of 0.1. } 

We initialized the temperature to be $10^4\,$K and calculated the initial pressure using an {adiabatic equation of state with $\gamma=5/3$}. Any variation in the initial pressure or temperature will be rapidly changed and dominated by the heating through the shock.

\subsection{The case of G1.9+0.3}

The high X-ray expansion speed and the absence {of any evidence for a central compact object or} a pulsar wind nebula are consistent with a type-Ia scenario for G1.9+0.3 \citep{2008ApJ...680L..41R}. We therefore applied the initial conditions suggested by \cite{1998ApJ...497..807D},
\begin{align}
 \rho_{\rm SN} = A\exp(-v/v_{\rm e})t_{\rm i}^{-3}\text{ and } v = r/t_{\rm i} \text{ with }\\
 v_{\rm e} = \left( \frac{E_{\rm ex}}{6M_{\rm ej}} \right)^{1/2} \text{ and } A =\frac{6^{3/2}}{8\pi}\frac{M_{\rm ej}^{5/2}}{E_{\rm ex}^{3/2}}\ .
\end{align}
Here $t_i$ is the time that has passed since the SN explosion, $M_{\rm ej}$ the ejecta mass, $E_{\rm ex}$ the explosion energy, and $r$ the spatial coordinate. 
For our simulations, we choose an initial age of about three months, for which the solution quickly converges against solutions with a lower initial age. {At that time the ejecta profile terminates at $\approx 10^{17}\,$cm, beyond which the gas density is that of the ISM. } The {cosmic-ray and turbulence calculations start with a 10-year old remnant, when the two-shock system is well established in the hydro profiles}. {We used approximately 500,000 grid points in our simulation domain that was initially spanning $3\,$pc}.

We used the canonical values for Type-1a explosions of $M_{\rm ej}=1.4M_\odot$ and $E_{\rm ex}=10^{51}\,$erg. The constant ambient density around the progenitor star then solely influences the evolution of the remnant. A value of $n=0.03\,\text{cm}^{-3}$ leads to a forward-shock radius of $R_{\rm fs}=2.2\,$pc at an age of $105\,$yrs with a forward shock speed of $v_{\rm fs}=14,000\,$km/s, which is in agreement with the extension of the bright X-ray ears measured with Chandra in 2008 \citep{2008ApJ...680L..41R}. The extension of the reverse shock is $R_{\rm rs}=1.85\,$pc at this age, which gives a ratio of $\frac{R_{\rm rs}}{R_{\rm fs}} = 0.84$. The reverse shock speed is $v_{\rm rs}=11000\,$km/s {in the observer frame and $5650\,$km/s in the plasma frame. At this early stage of evolution the density immediately downstream of the reverse shock is still three times higher than the density downstream of the forward shock. 

\begin{figure}[ht]
\includegraphics[width=0.48\textwidth]{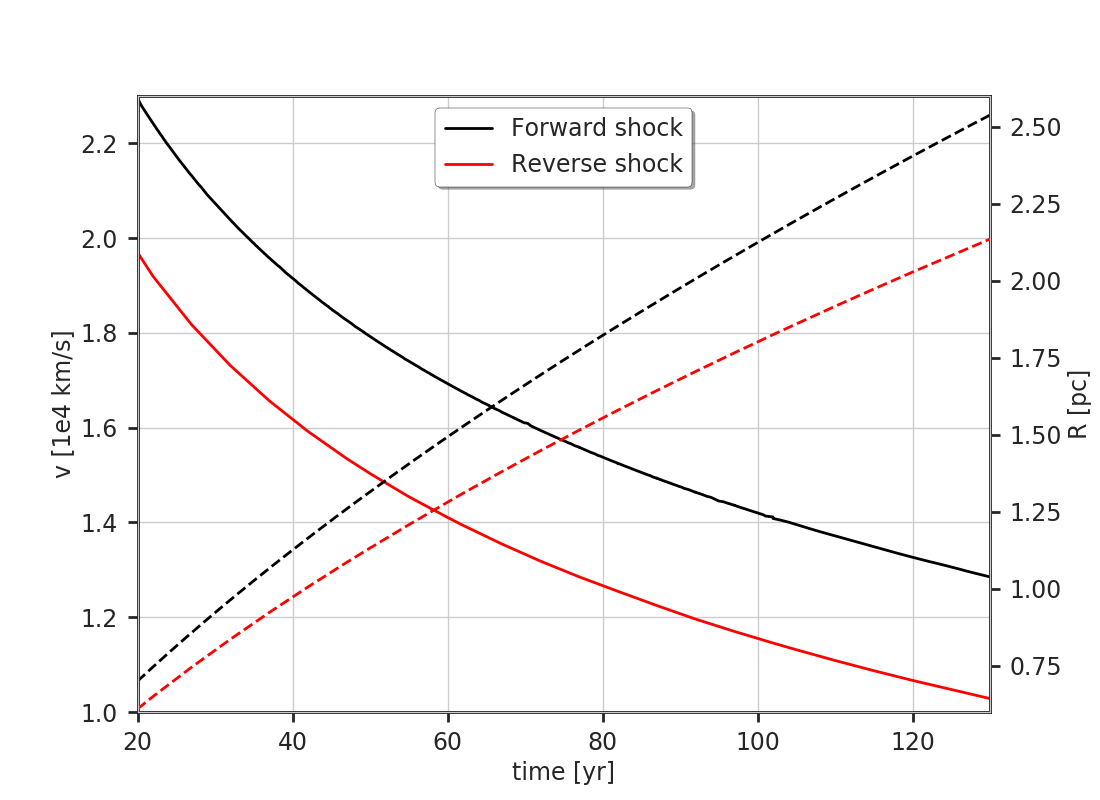}
\caption{Evolution of the {speed (solid lines) and radius (dashed lines) of the} forward and reverse shock in G1.9+0.3.}
\label{fig:speeds}
\end{figure} 
}

\section{Results}
We constructed two different models for the particle acceleration in G1.9+0.3. The first assumes that all the radio and X-ray emission of the SNR is produced only at the forward shock under the assumption of Bohm-like diffusion in an amplified field.
The second model takes the self-consistent amplification of turbulence at the forward shock into account and includes emission from the reverse shock.  

\subsection{One shock model}\label{sec:one-shock}
\begin{figure}[ht]
\includegraphics[width=0.48\textwidth]{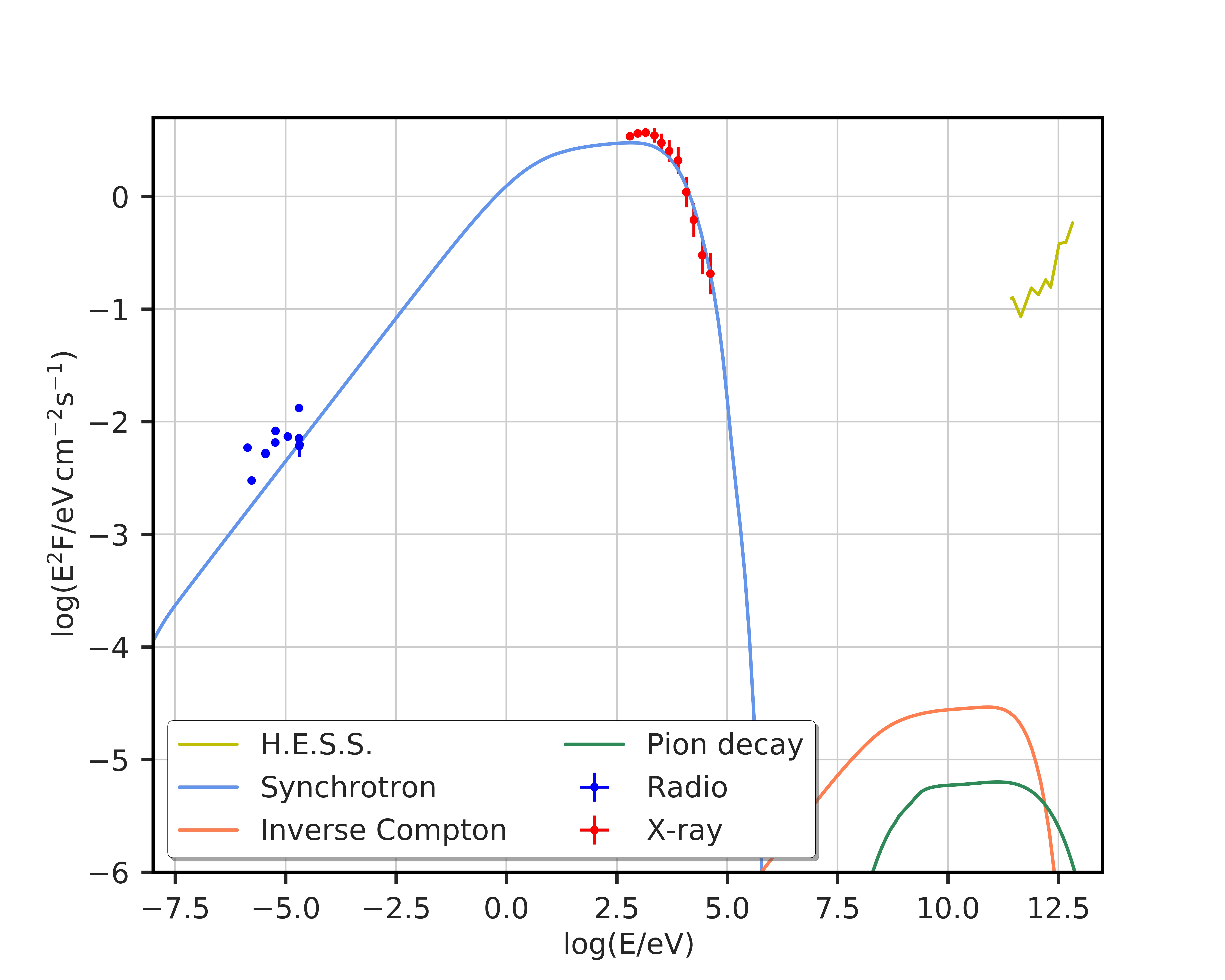}
\caption{Model SED for the forward shock only in a uniform upstream medium.}
\label{fig:SedOne_total}
\end{figure}

Figure \ref{fig:SedOne_total} shows the observed spectral energy distribution (SED) together with model curves representing the simulated emission from the remnant. Here we assumed Bohm-like diffusion in the downstream and in the upstream of the remnant up to a radius of $1.1\cdot R_\mathrm{sh}$. From $2\cdot R_\mathrm{sh}$ we used the galactic diffusion coefficient, and an exponential profile connects the two regimes in the intermediate range.  

To simultaneously reproduce the flux of radio and X-ray emission, a very strong magnetic field, $B_d\approx1.1\,$mG ($B_\mathrm{u}\approx330\,\mu$G), is needed to produce sufficient synchrotron cooling to significantly modify the electron spectrum at the young age of the remnant. 

To reproduce the observed synchrotron cutoff-energy, the diffusion coefficient must be increased by a factor of $30${, implying $\zeta=30$ in eq. (\ref{eq:Bohm}),} compared to Bohm-like diffusion, in agreement with \citet{2017A&A...603A...7A}, who concluded that the diffusion in the upstream of G1.9+0.3 can not be Bohm-like. Generally, Bohm-like diffusion is considered to be the most effective diffusion regime in SNRs, and the older remnants Cas A and RX J1713.4-3946 might well be within this regime \citep{2010ApJ...708..965Z,2014ApJ...785..130Z}. A deviation towards a less efficient diffusion regime might indicate slow growth of magnetic turbulence at the largest scales. 

Even though this model in principle explains the observed SED, the implied magnetic-field parameters are highly implausible. The downstream field corresponds to an energy density of
\begin{align}
 U_{\rm B,d} = \frac{B^2}{8\pi} = 3\cdot10^4\ \text{eV cm}^{-3}\text{,}
\end{align}
which is {a sizable fraction of} the energy density of the thermal plasma
\begin{align}
 U_{\rm Th,d} = \frac{3}{2}n_\mathrm{d}kT_\mathrm{d} = 10.8\cdot10^4\ \text{eV cm}^{-3}\text{.}
\end{align}
The energy density of CRs in the immediate downstream region is
\begin{align}
 U_{\rm CR} \approx \frac{c}{3}\int Np\ \text{d}p = 80\ \text{eV cm}^{-3}\text{.}
\end{align}
and considerably lower than that of the magnetic field.

Usually, just a few per cent of the thermal energy of the plasma are assumed to be transformed to magnetic-field energy - otherwise the evolution of SNRs should considerably deviate from purely hydrodynamical predictions - whereas here an $\beta \approx 0.3$, is needed. The magnetic-field value obtained in this way is also a factor 4 higher than that derived assuming equipartition with CRs, $\approx270\,\mu$G \citep{2014SerAJ.189...41D}. In the case of equipartition the energy density in CRs would be $\approx 1.8\cdot10^3 \text{eV cm}^{-3}$, which is about {$1.7\,$\%} of the energy density of the thermal plasma.

The strong synchrotron cooling also has effects on the morphology of the remnant. In the emission profiles, the X-ray peak is always close behind the forward shock, as only recently accelerated electrons have sufficient energy to emit X-ray photons. Radio photons, on the other hand, can be emitted by all previously accelerated electrons in the downstream, leading to an offset of about 6\% of the forward shock radius between the radio and the X-ray peaks in the emission profile. The observed separation between the peaks in the radio and the X-ray profiles is of the order $25\,$\% $R_\mathrm{sh}$ and hence incompatible with the one-shock scenario. None of the findings discussed above rely on a specific choice of turbulence modeling. They arise for both parametrized turbulence spectra and the solution of the transport equation for turbulence (cf. Eq.~\ref{Turb_1}). In the following, we shall explore the possibility of a two-shock scenario. 

\subsection{Two shock model}\label{sec:two-shock}
As the simple one-shock scenario fell short of explaining the morphology of the emission of G1.9+0.3 and raised further questions related to the apparently not Bohm-like diffusion regime, we build a second model taking the reverse-shock of G1.9+0.3 into account.

\begin{figure}[ht]
\includegraphics[width=0.49\textwidth]{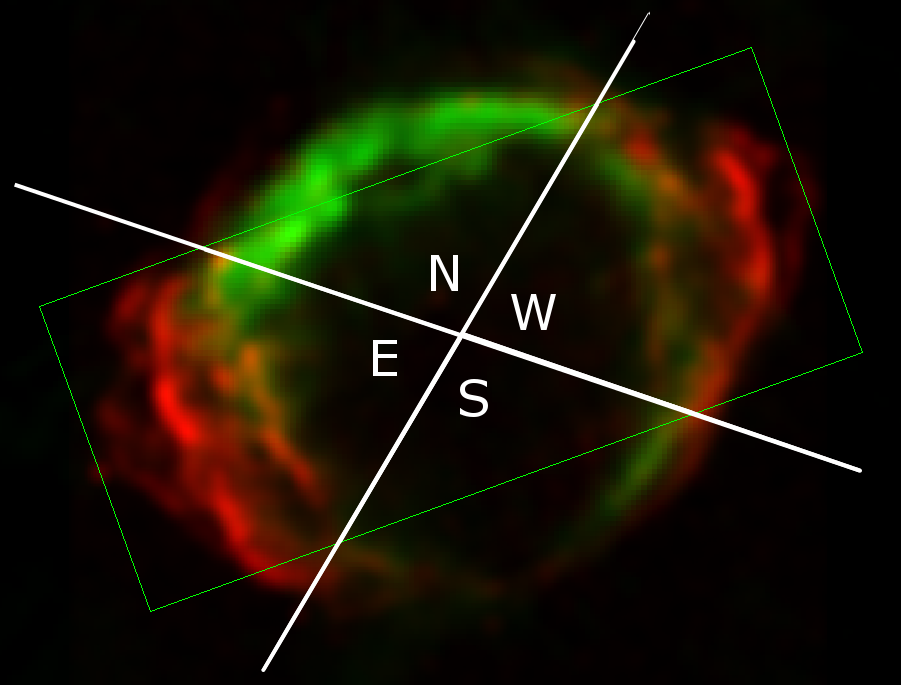}
\caption{Composite Image of X-ray (red, \cite{2017ApJ...837L...7B}) and radio (green, Luken et al. in prep.) observations of G1.9+0.3, both from 2017. The green box marks the region from which we extracted the emission profiles. The bulk of the X-ray emission is concentrated in the eastern (E) and western (W) cones. The radio emission is brightest in the north (N) and dimmer in the south (S).}
\label{fig:composite}
\end{figure} 

Figure \ref{fig:composite} illustrates the observed morphology of G1.9+0.3. We associate the X-ray emission with the forward shock because of its larger extension. The X-ray emission is concentrated in two cones in the east and the west of the remnant, while the radio emission is more spherically symmetric. Possible reasons for the concentration of the emission in the cones are discussed in section \ref{sec:morph}. The half-opening angle of the cones is $\approx 37.5^\circ$ in projection, which gives a volume filling factor  of about 20\% for the X-ray bright and thus acceleration-efficient regions of the forward-shock. As in a supersonic radial outflow there is little lateral transfer of energy and momentum, we can treat the cones as sections of a spherically symmetric model. Hence, we scale all emission from the forward shock in our spherical symmetric model with the factor 20\% to take into account the different radio and X-ray morphologies.

{A proper description of shock acceleration requires that the transport equation for CRs be calculated on a grid that provides very high resolution at the shock where particles are injected. Hence we need to solve it separately for particles injected at the forward and reverse shock, respectively. The transport equation for magnetic turbulence, which includes the contributions of the streaming instability and turbulent dynamo action at the shocks, requires knowledge of the CR distribution for the streaming-instability term. As we must treat CRs accelerated at the forward and reverse shock separately, the full CR distribution is available only a posteriori. Typically, the forward shock provides the dominant contribution to the streaming instability, and hence we calculate the streaming contribution for particle acceleration at the forward shock. Large-scale MHD turbulence arising from a turbulent dynamo is added both 
downstream of the forward and {downstream of} the reverse shock.}

The magnetic field {immediately behind} the forward shock is thus not an independent parameter as in the one-shock model and is determined by the injection parameter, $\xi$, {and the conversion efficiency of the flow energy to magnetic field in the downstream of the shock, $\kappa= 1.5\%$.} 

The streaming of particles from the forward shock only provides a very weak magnetic field at the reverse shock; its contribution to the total field strength is about 1\%. For particles accelerated at the reverse shock we assume {Bohm-like energy scaling for the diffusion coefficient for the particles accelerated at the reverse shock. The Larmor radius is calculated assuming that the total magnetic field near the reverse shock, that is obtained from the forward-shock calculation, can be interpreted as large-scale background magnetic field. This approximation appears justified for the low energy, and hence small Larmor radii, that particles accelerated at the reverse shock can attain. The scaling parameter for the diffusion coefficient, $\zeta$ in Eq.~\ref{eq:Bohm}, 
has with $\zeta=5000$ the highest value that still permits 
a synchrotron cutoff higher than typical radio frequencies.} The diffusion coefficient applied {upstream of} the reverse shock is about $1/275$ of the interstellar diffusion coefficient for particles of $1\,$GeV. 

\begin{figure}[ht]
\includegraphics[width=0.48\textwidth]{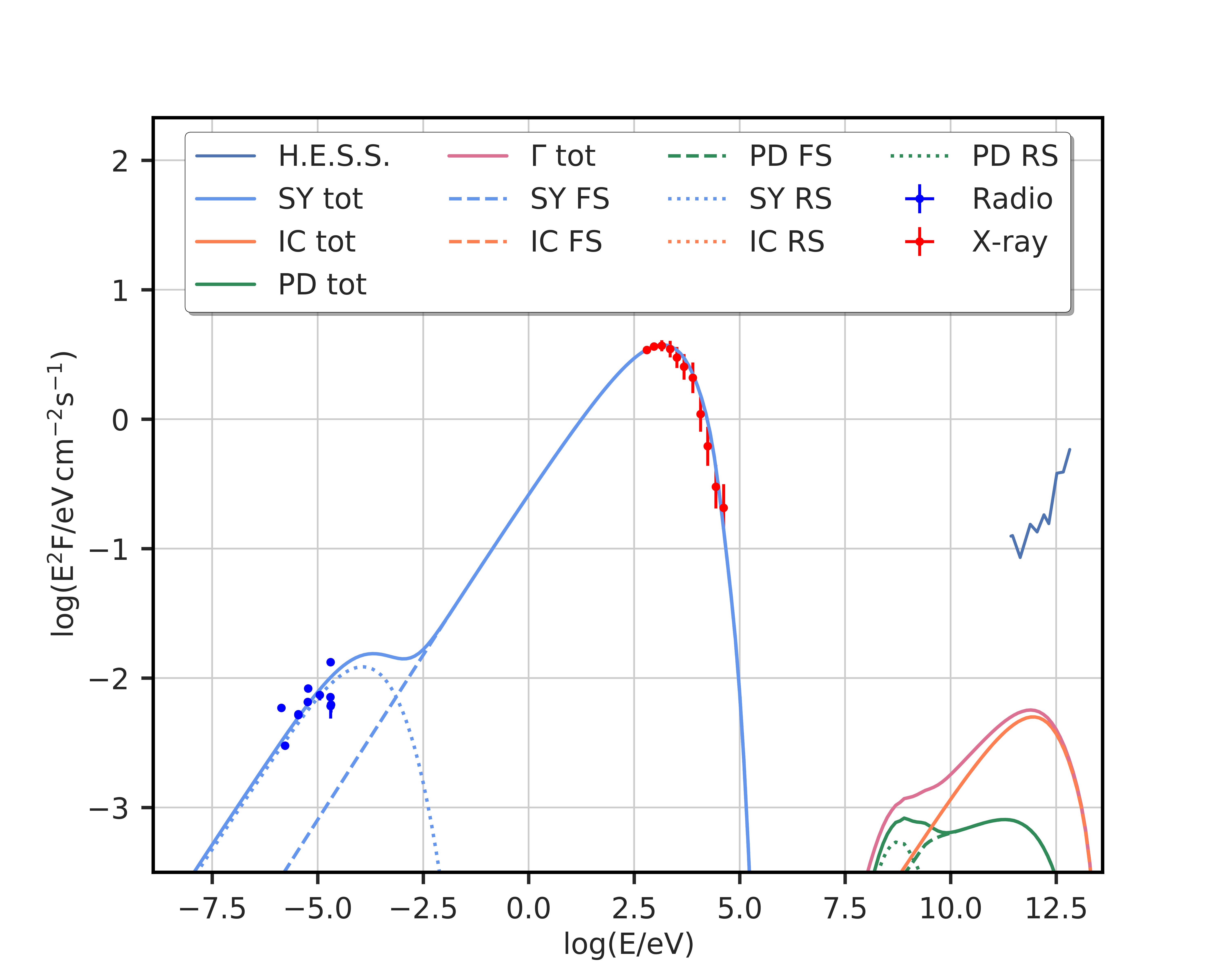}
\caption{Combined SED from the forward and reverse shock. FS (RS) denotes emission from accelerated at the forward (reverse) shock, and SY, PD, IC denote emission from synchrotron, pion decay, and inverse-Compton scattering, respectively.}
\label{fig:Sedtwo_total}
\end{figure}

The particle spectra obtained from both shocks are then combined to calculate the emission from the remnant. The results for this model are shown in Figure \ref{fig:Sedtwo_total}. 

We use an injection parameter of $\xi=3.65$ for the forward shock and a value of {$\xi=3.5$} for the reverse shock, which corresponds to {a three times higher probability of injection} there. The choice of different injection parameters is necessary to reproduce the measured flux-ratio of the radio and the X-ray emission. Another way of reproducing the observed emission would be to use different values for the injection efficiency of post-shock turbulence, $\kappa$, at the two shocks. 
The higher injection efficiency at the reverse shock might be attributed to the lower shock speed there, as it is easier for particles to catch up to the shock there. We assume an electron-to-proton ratio\footnote{The ratio is the relative injection efficiency of electrons to protons. A value of 1 results in a electron-to-proton ratio of $K_{\rm ep}=\sqrt{\frac{m_\mathrm{e}}{m_\mathrm{p}}}$ at high energies.} of 1/8 for both shocks and scale the emission of the forward-shock by 1/5 to account for the observed X-ray morphology, as discussed above.


The peak magnetic-field value behind the forward-shock is {180$\,\mu$G, and we find about 120$\,\mu$G downstream of the reverse shock. The {magnetic} energy density downstream of both shocks is thus only about $0.75\,$\% of the thermal energy density. The other $0.75\%$ of the thermal energy density that got converted into magnetic field are damped following cascading and converted back to thermal energy of the plasma.} 
For both shocks the ratio of CR pressure to shock ram pressure stays at $\approx 8.5\,$\%, justifying the test-particle assumption. {About $70\,\mu$G  from the $180\,\mu$G in the downstream are provided by the streaming of cosmic rays. This puts the ratio of $\delta B/B$ for the streaming instability to one, which is its supposed saturation level. 
}

\begin{figure}[ht]
\includegraphics[width=0.48\textwidth]{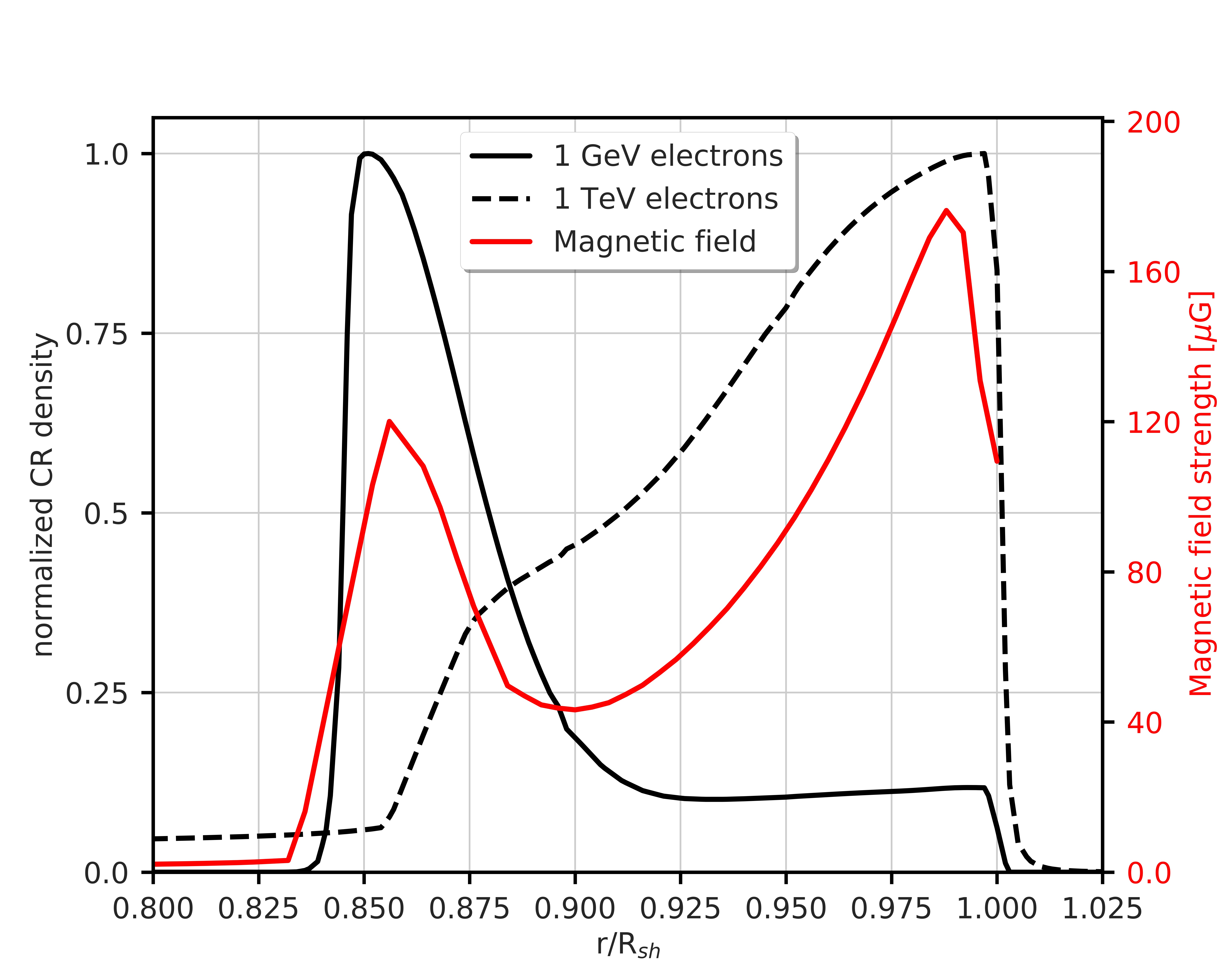}
\caption{Spatial distribution of the density of cosmic-ray electrons and the magnetic-field strength.}
\label{fig:CRandMFprofiles}
\end{figure}

In this model the radio emission is dominated by the reverse shock, whereas the X-ray synchrotron emission is dominated by the forward shock. The high plasma density at the reverse shock during the current stage of remnant evolution naturally explains the high number density of accelerated particles there. The relatively large diffusion coefficient at the reverse shock is still sufficient to accelerate CRs to energies high enough to explain the bulk of the radio emission of G1.9+0.3. The exact value of the diffusion coefficient at the reverse shock is not important, as long as the maximum frequency of synchrotron emission does not fall below the observed radio frequencies or the CR-pressure exceeds the test-particle limit. We illustrate the spatial distribution of high and low-energetic electrons together with the magnetic-field structure in figure \ref{fig:CRandMFprofiles}.

The magnetic-field profile we obtain strongly resembles the damped magnetic-field structure suggested by \cite{2005ApJ...626L.101P} and is shaped by the cascading of turbulence to smaller scales. Although no TeV-scale electrons are produced at the reverse shock, they can { diffuse there in the damped turbulence, and their number density is still 6\% of that at the forward shock}.

\subsubsection{Radio and X-ray Morphology}\label{sec:morph}

\begin{figure}[ht]
\includegraphics[width=0.48\textwidth]{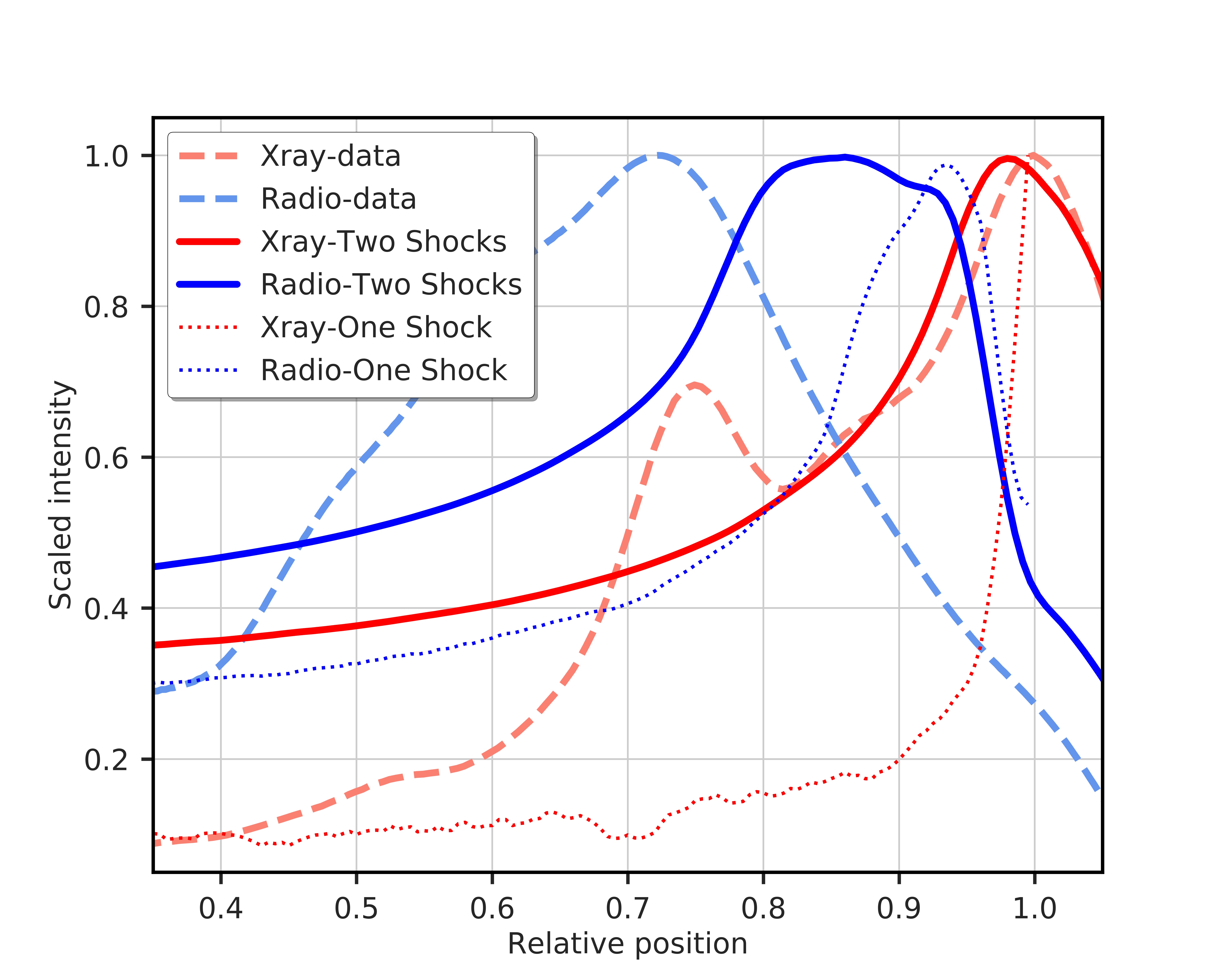}
\caption{Morphology of the remnant in X-rays and the radio band. The X-ray data are taken from \cite{2017ApJ...837L...7B} and the radio-data are from Luken et al. (in prep). The dashed lines indicate the intensity integrated in a rectangular region in the N-S direction covering the X-ray ears and projected onto the E-W-axis. A similar method was applied to the theoretical data. }
\label{fig:profiles}
\end{figure}

The strong radio emission of electrons accelerated at the reverse shock significantly increases the separation between the radio and the X-ray peak in the emission profiles. A comparison is shown in Figure \ref{fig:profiles}. The observed separation of the two peaks is about 25\% of the forward-shock radius, and accounting for two shocks increases the peak-separation in the model from 6.5\% in the one-shock scenario to about 15\%$\,R_\mathrm{sh}$ of the shock-radius.
The two-shock scenario may also explain the different morphology of the X-ray and the radio emission. The bipolar structure in X-rays is similar to the emission-structure in SN1006 and may reflect the orientation of the local magnetic field by either limiting the acceleration efficiency at a perpendicular shock or limiting the injection efficiency \citep{2003A&A...409..563V}. If so, the reverse shock would be unaffected by the magnetic field of the ISM. The magnetic-field structure expected in the unshocked ejecta would favor similar injection conditions over the entire reverse-shock surface.

The observed azimuthal variation in the intensity around the reverse-shock region can in this scenario still be attributed to a density gradient in the ISM with a higher density towards the north or, alternatively, an anisotropic explosion as indicated by the inhomogeneous distribution of heavier elements in the ejecta \citep{2013ApJ...771L...9B}. Both scenarios would lower(enhance) the expansion speed of the remnant towards the north(south) and lead to a higher(lower) density at the reverse shock. Thus the brightness difference between north and south would be attributed to the presence of more or less electrons that can be accelerated.

There are several potential explanations why in our model the intensity peaks in the radio and the X-ray band are closer together by $5$\%$\,R_\mathrm{sh}$ than they are observed to be. First, if the supernova explosion was anisotropic, our assumed ejecta distribution could be inappropriate or not be independent of direction. If the ejecta distribution were shallower, the position of the reverse shock would be further inward. A lower ejecta mass in one direction of the supernova expansion could result in faster deceleration of the forward shock with the consequence that the separation between forward and reverse shock would be larger.
Second, an asymmetric explosion or expansion in a inhomogeneous medium leads to a displacement of the geometric center of the remnant from the explosion site, and hence to a false estimate of the shock radius, yielding a different shock radius in each direction from the center of the explosion. It follows that the relative separation of the two shocks measured from the geometrical center of the explosion will be different in all directions. Modeling of an anisotropic explosion is beyond the scope of this paper, and we tried to minimize the impact of this distortion by averaging the E-W-profile in the N-S direction but as the exact geometry of the explosion is unknown, there might still be a remaining effect.
Third, as the emission of the forward and the reverse shock is decoupled, there is no reason why the effects determining the morphology of the X-ray - the ambient magnetic-field structure - and the radio emission - the density gradient or the structure of the explosion - should be aligned. If they are not aligned, then projection effects might result in a higher separation between the peaks of the X-ray and the radio emission. If, for example, the X-ray emission is aligned with the plane of the sky, then a tilt of the density gradient by $25\,^\circ$ is enough to shift the apparent position of the reverse shock from $84\,$\% of the forward shock radius to $75\,$\%.  

{It is important to note that the morphological difference between radio and X-ray emission makes G1.9+0.3 the first SNR {for which} non-thermal emission can be clearly attributed to the reverse shock. Non-thermal emission from the reverse shock has been discussed for RX J1713.7-3946 \citep{2010ApJ...708..965Z} and Cas A \citep{2014ApJ...785..130Z, 2008ApJ...686.1094H} as well but {on a much more tenuous basis than in the case of} G1.9+0.3. The model of  \cite{2010ApJ...708..965Z} for RX J1713.7-3946 requires a sizable non-thermal emission from the reverse shock only for a hadronic model of the gamma-ray emission of the remnant. If the gamma-ray emission is leptonic, an enhanced magnetic field towards the Rayleigh-Taylor unstable contact discontinuity can explain the observed radio and X-ray morphology of the remnant. However, the low density environment around RX J1713.7-3946 makes a leptonic scenario more likely unless a clumpy medium around the remnant provides enough target material \citep{2012ApJ...744...71I} {and there is a way to avoid emitting a detectable flux of X-rays \citep{2015A&A...577A..12F}}.
For Cas A \citet{2014ApJ...785..130Z} argue, that {a contribution from particles accelerated at the reverse shock is possible, but the radio morphology might also be attributed to an enhanced magnetic field towards the contact discontinuity. Although an association of X-ray emission with the reverse shock was suggested by earlier data \citep{2008ApJ...686.1094H}, the hard X-ray emission from Cas A appears to come from knots and filaments that are not well correlated with the reverse shock \citep{2015ApJ...802...15G}. However, these knots appear to be moving inwards far faster than expected for the reverse shock of Cas-A given its stage of evolution \citep{2018ApJ...853...46S}. }

The lack of X-ray emission from the position of the radio shell of G1.9+0.3 {is not in line with} an enhanced magnetic field near the contact discontinuity, because high-energy electrons accelerated at the forward shock should be able to diffuse there and produce intense X-ray emission which is not observed. }

\subsubsection{Radio Brightening}
The large dataset {collected} by MOST shows a brightening of the total radio flux at 843$\,$MHz rate at the rate {of} {$1.22^{+0.24}_{-0.16}\,$\%/yr} \citep{2008MNRAS.389L..23M}. {This brightening can be attributed either to an increase of the magnetic field strength or to an increase of the number of radiating electrons. }

{For a type-Ia supernova the relative increase in the number of radiating particles}
can be estimated as
\begin{align}
 \frac{d V}{V} &=3 \frac{dr}{r}=3\frac{v dt}{r}\text{ , }
\end{align}
{resulting in} 
$1.9\,$\%/yr for the forward shock and $1.6\,$\%/yr for the reverse shock.

In our simulations {we obtain a value of $0.75\,$\%/yr,} which is compatible {within $~2\sigma$ with the observed rate } {but lower than the brightening expected for a constant magnetic-field strength and a constant upstream density. However, the density upstream of the reverse shock is decreasing with a rate of $\approx 0.7\,$\%/yr which reduces the expected analytic brightening rate to $\approx 0.9\,$\%/yr.} 

\begin{figure}[ht]
\includegraphics[width=0.49\textwidth]{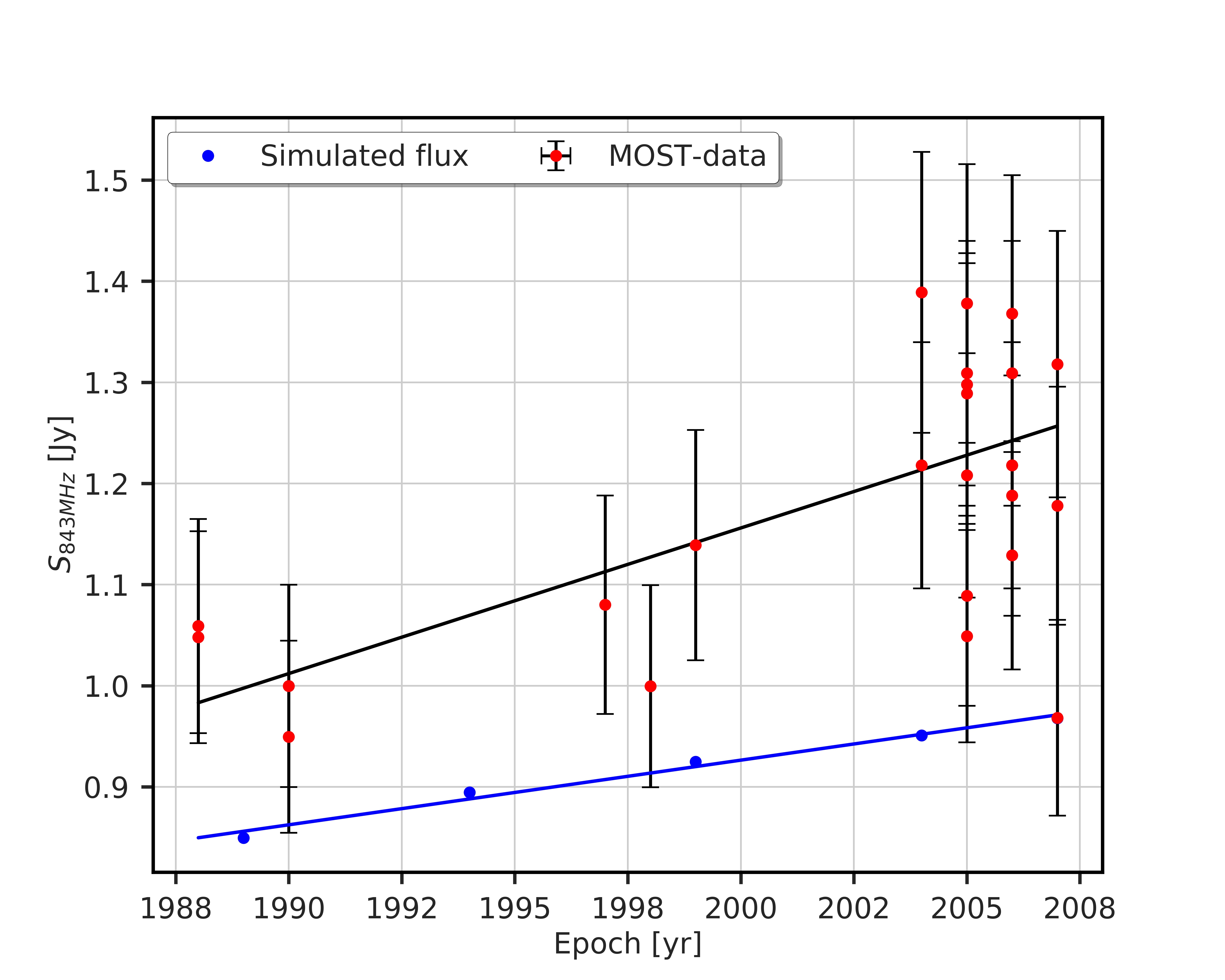}
\caption{Time evolution of the radio flux at 843MHz.}
\label{fig:Brghtening}
\end{figure} 

{Alternatively,} if the evolution of the number of electrons is well known, 
the {observed} brightening rate {can} provide a measurement of the change of the magnetic-field {strength} in the remnant. The observed value of about $1.2\,$\%/yr {would be reproduced, if the magnetic-field strength} {increased} by $0.15\,$\%/yr. {At this time, the uncertainty in the rate of radio brightening is too large to establish a non-zero evolution of the magnetic-field strength.}  

None of the observed properties of G1.9+0.3, except the radio brightening, depends on our parametrization of the post-shock turbulence. A significant fraction of the magnetic field is assumed to arise from MHD instabilities at the shock. If we coupled the injection of turbulence dynamically to the energy density in the post-shock flow, the magnetic field would decrease over time as the shock velocity decreases at both shocks and so does the plasma density at the reverse shock. Taking this at face value, the radio brightness of G1.9+0.3 would decrease by $0.23\,$\%/yr. The data suggest otherwise, and so we implemented a fixed injection of large-scale turbulence, leading to an almost constant magnetic field. In reality, part of the MHD turbulence may be triggered by cosmic-ray interactions in the precursor that need time to build up \citep{2009ApJ...707.1541B,2016MNRAS.458.1645D}, and the efficiency of MHD-turbulence injection may have a non-trivial time profile.


As the X-ray emission originates at the forward shock, we expect its brightening rate to be different from that of the radio emission. Our model yields a brightening rate of $1.5\,$\%/yr, which is {sligthly below} the analytic value we expect for a constant magnetic field and close to {the observed one}, $1.9\pm0.4\,$\%/yr \citep{2014ApJ...790L..18B} and $1.3\pm0.8\,$\%/yr \citep{2017ApJ...837L...7B}.



We did not account for an amplification of magnetic field at the reverse shock through, e.g., processes driven by accelerated cosmic rays or Rayleigh-Taylor and Helmholtz instabilities that will develop at the contact discontinuity. This would provide additional magnetic field at or close to the reverse shock and allow to fit the SED with fewer particles injected at the reverse shock. Likewise, the structure of the medium encountered by the shock is decisive for the development of post-shock turbulence. For example, an inhomogeneous distribution of the ejecta may increase the conversion efficiency of flow energy to magnetic field at the reverse shock. Our coupling of the energy density transferred to post-shock turbulence to the energy density of the flow is a likely over-simplified parametrization of the physics at play. 

\subsubsection{Prospects for CTA}

\begin{figure}[ht]
\includegraphics[width=0.48\textwidth]{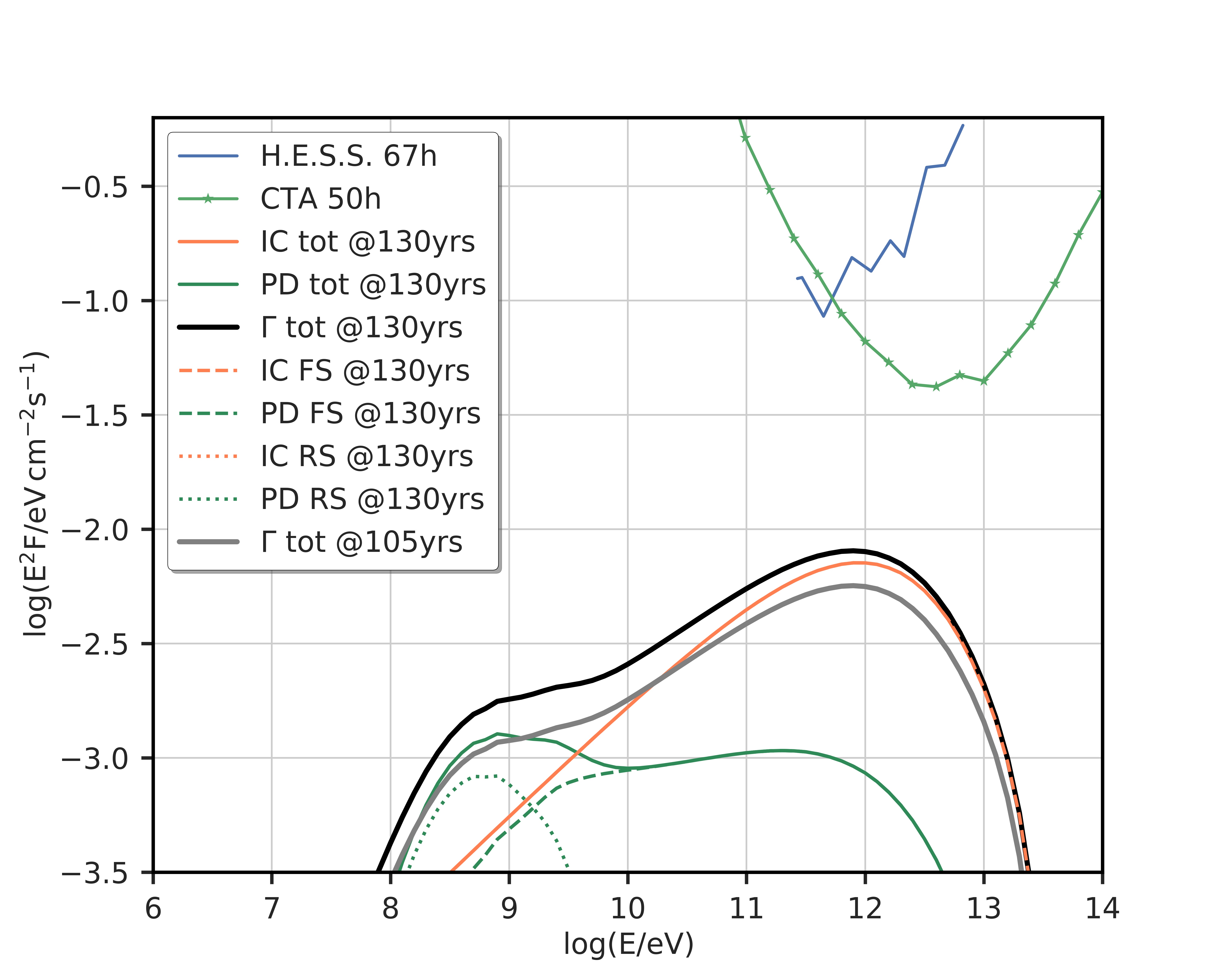}
\caption{Gamma-ray spectrum from G1.9+0.3 in 2007 (grey line) and 2032 (black line). IC, PD denote emission from inverse-Compton scattering and pion decay respectively. FS (RS) denotes emission from the forward (reverse) shock.}
\label{fig:CTA}
\end{figure}

The TeV-band gamma-ray emission from G1.9+0.3 is dominated by inverse Compton scattering of CMB and IR photons off high-energy electrons accelerated at the forward shock. {We approximated the IR-radiation field from \cite{2006ApJ...648L..29P} at the location of G1.9+0.3 with a {grey-body distribution with temperature $T_{IR}=55\,$K and energy density} $U_{IR}=1.5\,$eV/cm$^3$.} Figure (\ref{fig:CTA}) shows that our predictions for the gamma-ray flux are well in agreement with the H.E.S.S upper limits and about {six} times below the {design} sensitivity of CTA south for 50h of observations.

{Given that G1.9+0.3 is located within the survey field of the Galactic center with more than 500 hours of observations planned for that region, a $3\sigma$ detection of the signal from G1.9+0.3 could still be possible in the event of future improvements in sensitivity over the  original design specifications.}


We expect the gamma-ray flux to increase by about 25\% in the epoch between 2007 and 2032. The gamma-ray luminosity of G1.9+0.3 above $1\,$TeV would be $\approx1.0\cdot10^{32}\,$erg/s, which is well within the range found by H.E.S.S. in their SNR population study \citep{2018A&A...612A...3H}. 

\section{Conclusions}\label{sec:conc}

We modeled the broadband non-thermal emission from the supernova remnant G1.9+0.3 with a view to analyze the observed spectral energy distribution, the morphology, the brightening rate at radio frequencies and to elucidate the prospects for detecting the remnant with CTA. For that purpose we solved the transport equations for cosmic rays and Alfv\'enic turbulence together with the standard gas-dynamical equations, and we subsequently calculated the emission spectra from the forward and the reverse shock under the assumption of spherical symmetry.

We showed that a scenario involving only the forward shock needs a very strong magnetic field in order to explain the SED of the remnant and falls short of explaining the separation between the peaks of radio and X-ray emission. The required magnetic-field strength in this scenario is well above the equipartition value and implies the same energy density in the magnetic field and in the thermal plasma, which is very unlikely.

Allowing for electron acceleration at the reverse shock reproduces the SED with lower magnetic-field strength and decouples the X-ray and radio emission, the former arising from multi-TeV electrons accelerated at the forward shock and the latter produced by the large number of lower-energy electrons energized at the reverse shock. While the bipolar X-ray morphology might be the result of a variation in acceleration efficiency due to the magnetic-field structure, the radio morphology would be unaffected by this. A density gradient in the ambient medium towards the north or an asymmetric explosion is sufficient to explain the radio N-S asymmetry. It would also account for the difference in expansion velocity measured between X-ray and radio features in the E-W direction and better reproduce the separation between the radio and X-ray peaks.

{The observed radio brightening rate is consistent with a magnetic field that is constant over time. Our model predicts a brightening rate of $0.75\,$\%/yr, compatible within $2\rb{\sigma}$ with the observed $1.22^{+0.24}_{-0.16}\,$\%/yr.}

The limited time available to drive large-scale turbulence in the upstream region naturally causes super-Bohmian diffusion and hence an acceleration efficiency well below that expected for Bohm diffusion. The maximum energy of particles is below 100 TeV, and so G1.9+0.3 is not a PeVatron.

We further evaluated the prospects of detecting G1.9+0.3 with the CTA observatory. Our model predicts a brightening of G1.9+0.3 by $\approx25\,$\% between 2007 and 2032{ and a gamma-ray flux that is approximately six times below the design sensitivity of CTA for 50h of observations.}. Essentially all of the TeV-scale gamma-ray flux is produced by particles accelerated at the forward shock, and the majority comes from inverse-Compton scattering. Pion-decay radiation contributes about 20\% of the flux.

Our model indicates that G1.9+0.3 is the first SNR with detected non-thermal emission that can be {clearly} attributed to the reverse shock.


\bibliographystyle{aa}
\bibliography{References}

\end{document}